\documentclass[a4paper,10pt]{article}

\usepackage{amsmath,amssymb,mathtools}     
\usepackage{color}
\usepackage{graphicx}
\usepackage{subfigure}
\usepackage{cite}                
\usepackage{hyperref}            
\usepackage{multirow,makecell}   
\usepackage{textcomp}
\usepackage{wasysym}

\usepackage[text={17.2cm,25.2cm},centering]{geometry}

\numberwithin{equation}{section}   

\def \be {\begin{equation}}
\def \ee {\end{equation}}
\def \ba {\begin{array}}
\def \ea {\end{array}}
\def \bea{\begin{eqnarray}}
\def \eea{\end{eqnarray}}
\def \nn {\nonumber}

\def \a {\alpha}
\def \b {\beta}

\def \d {\delta}

\def \e {\epsilon}

\def \L {\Lambda}

\def \r {\rho}

\def \mA {\mathcal A}
\def \mB {\mathcal B}
\def \mC {\mathcal C}
\def \mD {\mathcal D}
\def \mE {\mathcal E}
\def \mF {\mathcal F}

\def \mI {\mathcal I}

\def \mK {\mathcal K}
\def \mL {\mathcal L}
\def \mM {\mathcal M}
\def \mN {\mathcal N}
\def \mO {\mathcal O}
\def \mP {\mathcal P}

\def \mX {\mathcal X}

\def \p {\partial}
\def \f {\frac}

\def \mc {\mathcal}

\def \lt {\left}
\def \rt {\right}

\def \inf {\infty}

\def \lag {\langle}
\def \rag {\rangle}

\def \ep {\mathrm{e}}
\def \ii {\mathrm{i}}

\def \tr {\textrm{tr}}

\def \and {{\textrm{and}}}

\def \Z {{\textrm{Z}}}

\def \vac {{\langle0|}}
\def \uum {{|0\rangle}}

\def \vacc {{\textrm{vac}}}

\def \cl {{\textrm{cl}}}

\def \oloop {{\textrm{1-loop}}}

\def \cL {{\textrm{L}}}
\def \NL {{\textrm{NL}}}
\def \NNL {{\textrm{NNL}}}

\begin{document}

\title{\textbf{Holographic R\'enyi entropy for two-dimensional $\mathcal{N}$=(2,2) superconformal field theory}}
\author{Zhibin Li$^{1,2}$\footnote{lizb@ihep.ac.cn}\,
and Jia-ju Zhang$^3$\footnote{jiajuzhang@outlook.com}}
\date{}

\maketitle

\vspace{-10mm}

\begin{center}
{\it
$^{1}$Theoretical Physics Division, Institute of High Energy Physics, Chinese Academy of Sciences,\\
19B Yuquan Rd, Beijing 100049, P.R.\,China\\ \vspace{1mm}

$^{2}$Theoretical Physics Center for Science Facilities, Chinese Academy of Sciences,\\
19B Yuquan Rd, Beijing 100049, P.R.\,China\\ \vspace{1mm}

$^{3}$ Dipartimento di Fisica, Universit\'a degli studi di Milano-Bicocca, Piazza della Scienza 3, I-20126 Milano, Italy
}
\vspace{10mm}
\end{center}

\begin{abstract}

  We investigate the holographic R\'enyi entropy for two-dimensional $\mathcal N=(2,2)$ superconformal field theory (SCFT), which is dual to $\mathcal N=2$ supergravity in AdS$_3$ background. In SCFT we have the stress tensor, current, and their supersymmetric partners, and in supergravity we have the graviton, vector field, and two gravitinos. We get the R\'enyi mutual information of two short intervals on complex plane in expansion by the cross ratio $x$ to order $x^4$, and R\'enyi entropy of one interval on torus in expansion by $q=\exp(-2\pi\beta/L)$, with $\beta$ being the inverse temperature and $L$ being the spatial period, to order $q^2$. We calculate in both the supergravity and SCFT sides, and find matches of the results.

\end{abstract}

\baselineskip 18pt
\thispagestyle{empty}
\newpage

\tableofcontents

\section{Introduction}

Holographic entanglement entropy \cite{Ryu:2006bv,Ryu:2006ef} is an interesting subject in the investigation of AdS/CFT (anti-de Sitter/conformal field theory) correspondence \cite{Maldacena:1997re,Gubser:1998bc,Witten:1998qj,Aharony:1999ti}. To calculate the entanglement entropy one can use the replica trick, calculate the $n$-th R\'enyi entropy with $n \in \Z$ and $n>1$, and take the $n\to1$ limit to get the entanglement entropy \cite{Callan:1994py,Holzhey:1994we,Calabrese:2004eu}. R\'enyi entropy is not only a tool to calculate the entanglement entropy but also interesting in its own right, and one can directly investigate the holographic R\'enyi entropy \cite{Headrick:2010zt,Hung:2011nu,Fursaev:2012mp,Hartman:2013mia,Faulkner:2013yia,Barrella:2013wja,Dong:2016fnf}.

Quantum gravity in AdS$_3$ background with cosmological constant $\L=-\f{1}{R^2}$ and small Newton constant $G_N$ is dual to a two-dimensional CFT with a large central charge \cite{Brown:1986nw}
\be
c=\f{3R}{2G_N}.
\ee
There are many investigations of holographic R\'enyi entropy in AdS$_3$/CFT$_2$ correspondence
\cite{Headrick:2010zt,Calabrese:2010he,Hartman:2013mia,Faulkner:2013yia,Barrella:2013wja,Chen:2013kpa,%
Chen:2013dxa,Perlmutter:2013paa,Chen:2014kja,Beccaria:2014lqa,Cardy:2014jwa,Chen:2014unl,Long:2014oxa,%
Chen:2015kua,Chen:2015uia,Chen:2015uga,Zhang:2015hoa,Li:2016pwu,Chen:2016uvu,Chen:2016lbu,Zhang:2016rja}.
In gravity side, one can not only get the classical part \cite{Faulkner:2013yia,Barrella:2013wja,Chen:2014unl} but also use the partition function of Einstein gravity in handlebody background \cite{Maloney:2007ud,Yin:2007gv,Giombi:2008vd} and get the one-loop corrections \cite{Barrella:2013wja}. In CFT side one can use the operator product expansion (OPE) of twist operators to calculate the R\'enyi mutual information of two short intervals on complex plane \cite{Headrick:2010zt,Calabrese:2010he,Hartman:2013mia,Chen:2013kpa}, and use the low temperature expansion of density matrix to calculate the R\'enyi entropy of one interval on torus with low temperature \cite{Cardy:2014jwa,Chen:2014unl}.

In this paper we investigate the holographic R\'enyi entropy in another kind of AdS$_3$/CFT$_2$ correspondence, the dual between the $\mN=2$ supergravity in AdS$_3$ background and the $\mN=(2,2)$ superconformal field theory (SCFT). In supergravity side we have the fields of graviton, vector and two gravitinos, and we denote them by $g$, $v$ and $\psi$, respectively. Correspondingly, in SCFT side we have the stress tensor $T$, $\bar T$, current $J$, $\bar J$, and their superpartners $G$, $\bar G$, $H$, $\bar H$. Note that the duality between the bosonic part of the supergravity with only graviton and vector field and the bosonic part of the SCFT with only stress tensor and current is self-consistent, and we will also consider the holographic R\'enyi entropy in such a correspondence.

The remaining part of this paper is arranged as follows. In section~\ref{sec2}, we investigate the R\'enyi mutual information of two short intervals on complex plane. In section~\ref{sec3}, we consider the R\'enyi entropy of one interval on tours with low temperature. Then there is the conclusion and discussion in section~\ref{sec4}. In appendix~\ref{appA}, we review some details of the $\mN=(2,2)$ SCFT. In appendix~\ref{appB}, we give some useful summation formulas.

\section{R\'enyi mutual information of two intervals on complex plane} \label{sec2}

In this section we calculate the R\'enyi mutual information of two short intervals on complex plane. In other words, we have the ground state SCFT on space of one-dimensional infinite straight line. We expand the R\'enyi mutual information by the small cross ratio $x$ to order $x^4$. In gravity side we use method in \cite{Barrella:2013wja}, and in CFT side we use OPE of twist operators in \cite{Headrick:2010zt,Calabrese:2010he,Chen:2013kpa}.

\subsection{Supergravity result}

The classical part of the holographic R\'enyi mutual information only depends on the classical configuration of the gravity background, and so it is the same as that of pure Einstein gravity. The result can be found in \cite{Barrella:2013wja}
\be \label{Incl}
I^{\cl}_n = \frac{c (n-1) (n+1)^2 x^2}{144 n^3}+\frac{c (n-1) (n+1)^2 x^3}{144 n^3}
           +\frac{c (n-1) (n+1)^2 (1309 n^4-2 n^2-11) x^4}{207360 n^7} + O(x^5).
\ee
To calculate the one-loop part of the holographic R\'enyi mutual information to order $x^4$, we need the contributions of the consecutively decreasing words (CDWs) and the 2-CDWs \cite{Barrella:2013wja}. The one-loop part depends on the field content of the gravity theory. For the graviton $g$ we have \cite{Barrella:2013wja}
\be \label{Ingoloop}
I_{n,g}^\oloop = \frac{(n+1) (n^2+11) (3 n^4+10 n^2+227) x^4}{3628800 n^7}+O(x^5),
\ee
for the vector field $v$ we have
\bea \label{Involoop}
&& I_{n,v}^\oloop = \frac{(n+1) (n^2+11) x^2}{720 n^3}+\frac{(n+1) (23 n^4+233 n^2-40) x^3}{15120 n^5}\\
&& \phantom{I_{n,v}^\oloop =}
                   +\frac{(n+1) (2129 n^6+20469 n^4-6285 n^2+2695) x^4}{1451520 n^7}+O(x^5),\nn
\eea
and for the two gravitinos $\psi$ we have
\be \label{Inpsioloop}
I_{n,\psi}^\oloop = \frac{(n+1) (2 n^4+23 n^2+191) x^3}{30240 n^5}
                   +\frac{(n+1) (33 n^6+358 n^4+2857 n^2-368) x^4}{302400 n^7} + O(x^5).
\ee

\subsection{Contributions of $T$ in SCFT}

Using the replica trick we get the $n$-fold of the original SCFT, which we call SCFT$^n$. In SCFT$^n$ we use OPE of twist operators to calculate the partition function, and the R\'enyi mutual information turns out to be \cite{Chen:2013kpa,Chen:2013dxa,Perlmutter:2013paa}
\be \label{In}
I_n=\f{2}{n-1}\log\Big[ \sum_K \a_K d_K^2 x^{h_K} {}_2F_1(h_K,h_K;2h_K;x) \Big],
\ee
where the summation of $K$ is over all the linearly independent orthogonalized holomorphic quasiprimary operators $\Phi_K$ in SCFT$^n$. Here $\a_K$, $d_K$, $h_K$ are, respectively, the normalization factor, OPE coefficient, and conformal weight of $\Phi_K$. Note that there is a factor 2 in the right-hand side of (\ref{In}), and this concerns the contributions from the anti-holomorphic sector.

In this subsection we repeat the calculation in \cite{Chen:2013kpa}, with the contributions of only the vacuum conformal family, i.e.\ the contributions of operators that are constructed solely by $T$. To level 4 the SCFT$^n$ quasiprimary operators we have to consider are counted as
\be \label{z19}
(1-x)\chi_{(0)}^n+x = 1+n x^2+\frac{n (n+1)}{2} x^4+O(x^5),
\ee
with $\chi_{(0)}$ in (\ref{chi0}), and they are listed in table~\ref{qpnT}. For these quasiprimary operators we have the normalization factors
\be
\a_1=1, ~~ \a_T=\f{c}{2}, ~~ \a_\mA = \f{c(5c+22)}{10}, ~~ \a_{TT} = \f{c^2}{4},
\ee
and the OPE coefficients
\be
d_1=1, ~~
d_T=\f{n^2-1}{12n^2}, ~~
d_\mA=\f{(n^2-1)^2}{288n^4}, ~~
d_{TT}^{j_1j_2}=\f{1}{8n^4c}\f{1}{s_{j_1j_2}^4}+\f{(n^2-1)^2}{144n^4}.
\ee
Using (\ref{In}) we get the R\'enyi mutual information organized by the large central charge
\be
I_n^{(0)} = I_{n}^\cL + I_{n,(0)}^\NL + \cdots,
\ee
with the leading part $I_{n}^\cL$ equaling $I_n^\cl$ (\ref{Incl}) and the next-to-leading part $I_{n,(0)}^\NL$ equaling $I_{n,g}^\oloop$ (\ref{Ingoloop}).

\begin{table}\centering
\begin{tabular}{|c|c|c|c|}\hline
  level            & operator & degeneracy      & number                         \\ \hline
  0                & 1        & 1               & 1                              \\ \hline
  2                & $T$      & $n$             & $n$                            \\ \hline
  \multirow{2}*{4} & $\mA$    & $n$             & \multirow{2}*{$\f{n(n+1)}{2}$} \\ \cline{2-3}
                   & $TT$     & $\f{n(n-1)}{2}$ &                                \\ \hline
\end{tabular}
\caption{To level 4 the holomorphic quasiprimary operators in SCFT$^n$ with contributions of only $T$. We follow the convention in \cite{Li:2016pwu} and omit the replica indices that take values from 0 to $n-1$. Note that the 4th column agrees with the counting in (\ref{z19}).} \label{qpnT}
\end{table}

\subsection{Contributions of $T$, $J$ in SCFT}

\begin{table}\centering
\begin{tabular}{|c|c|c|c|c|}\cline{1-5}
  level & operators & ? & degeneracy & number \\ \cline{1-5}

  1 & $J$ & \texttimes & $n$ & $n$ \\ \cline{1-5}

  \multirow{2}*{2} & $\mM$ & \texttimes & $n$             & \multirow{2}*{$\f{n(n+1)}{2}$} \\ \cline{2-4}
                   & $JJ$  & \checked   & $\f{n(n-1)}{2}$ & \\ \cline{1-5}

                   & $\mK,~\mO$ & \texttimes & $2n$                 & \\ \cline{2-4}
  \multirow{2}*{3} & $TJ,~J\mM$ & \texttimes & $2n(n-1)$            & \multirow{2}*{$\f{n (n^2+12 n-1)}{6}$} \\ \cline{2-4}
                   & $JJJ$      & \texttimes & $\f{n(n-1)(n-2)}{6}$ & \\ \cline{2-4}
                   & $I(JJ)$    & \checked   & $\f{n(n-1)}{2}$      & \\ \cline{1-5}

                   & $\mL,~\mN,~\mP_1,~\mP_2$ & \texttimes & $4n$                       & \\ \cline{2-4}
                   & $T\mM,~J\mO$             & \texttimes & $2n(n-1)$                  & \\ \cline{2-4}
                   & $J\mK,~\mM\mM$           & \checked   & $\f{3n(n-1)}{2}$           & \\ \cline{2-4}
  \multirow{2}*{4} & $TJJ,~JJ\mM$             & \checked   & $n(n-1)(n-2)$              & \multirow{2}*{$\frac{n(n^3+26n^2+59n+10)}{24}$} \\ \cline{2-4}
                   & $JJJJ$                   & \checked   & $\f{n(n-1)(n-2)(n-3)}{24}$ & \\ \cline{2-4}
                   & $I(TJ),~I(J\mM)$         & \texttimes & $2n(n-1)$                  & \\ \cline{2-4}
                   & $I(JJJ)$                 & \texttimes & $\f{n(n-1)(n-2)}{3}$       & \\ \cline{2-4}
                   & $II(JJ)$                 & \checked   & $\f{n(n-1)}{2}$            & \\ \cline{1-5}
\end{tabular}
\caption{Extra holomorphic quasiprimary operators in SCFT$^n$ after adding contributions of $J$. As in \cite{Li:2016pwu}, we use the notations $I(JJ)=J\ii\p J- \ii\p JJ$, $II(JJ) = \p J\p J-\f13 (J\p^2J+\p^2JJ)$, and the ones similar to them. In the 3rd column we use \checked\ or \texttimes\ to denote whether the operators contribute to the R\'enyi mutual information or not. The 5th column agrees with the counting in (\ref{z20}).}
\label{qpnJ}
\end{table}

By adding the contributions of $J$, we have to consider the extra quasiprimary operators that are counted as
\be \label{z20}
(1-x)\chi_{(0)}^n(\chi_{(1)}^n-1) =  n x+\frac{n (n+1)}{2} x^2+\frac{n (n^2+12n-1)}{6} x^3
                                   + \frac{n(n^3+26n^2+59n+10)}{24} x^4 + O(x^5),
\ee
with $\chi_{(0)}$ being (\ref{chi0}) and $\chi_{(1)}$ being (\ref{chi1}), and they are listed in table~\ref{qpnJ}. Only a few of them contribute to the R\'enyi mutual information. For the relevant operators we have the normalization factors
\bea
&& \a_{JJ}=\f{c^2}{9}, ~~ \a_{I(JJ)}=\f{4c^2}{9}, ~~ \a_{J\mK}=\f{c^2(c+2)}{18}, ~~ \a_{\mM\mM}=\f{4c^2(c-1)^2}{81}, \nn\\
&& \a_{TJJ}=\f{c^3}{18}, ~~ \a_{JJ\mM}=\f{2c^3(c-1)}{81}, ~~ \a_{JJJJ}=\f{c^4}{81}, ~~ \a_{II(JJ)}=\f{20c^2}{27},
\eea
and the OPE coefficients
\bea
&& \hspace{-8mm}
   d_{JJ}^{j_1j_2} = -\f{3}{4n^2c}\f{1}{s_{j_1j_2}^2}, ~~
   d_{I(JJ)}^{j_1j_2} = -\f{3}{8n^3c}\f{c_{j_1j_2}}{s_{j_1j_2}^3}, ~~
   d_{J\mK}^{j_1j_2} = -\f{n^2-1}{16n^4c}\f{1}{s_{j_1j_2}^2}, ~~
   d_{\mM\mM}^{j_1j_2} = \f{9}{32n^4c(c-1)}\f{1}{s_{j_1j_2}^4},   \nn\\
&& \hspace{-8mm}
   d_{TJJ}^{j_1j_2j_3} = \f{1}{16n^4} \Big( \f{6}{c^2}\f{1}{s_{j_1j_2}^2 s_{j_1j_3}^2}-\f{n^2-1}{c}\f{1}{s_{j_2j_3}^2} \Big), ~~
   d_{JJ\mM}^{j_1j_2j_3} = \f{9}{16n^4c^2}\f{1}{s_{j_1j_3}^2 s_{j_2j_3}^2},  \\
&& \hspace{-8mm}
   d_{JJJJ}^{j_1j_2j_3j_4} = \f{9}{16n^4c^2} \Big(   \f{1}{s_{j_1j_2}^2 s_{j_3j_4}^2}
                                                   + \f{1}{s_{j_1j_3}^2 s_{j_2j_4}^2}
                                                   + \f{1}{s_{j_1j_4}^2 s_{j_2j_3}^2} \Big), ~~
  d_{II(JJ)}^{j_1j_2} = -\f{3}{160n^4c} \Big( \f{15}{s_{j_1j_2}^4} - \f{2(n^2+5)}{s_{j_1j_2}^2} \Big), \nn
\eea
with the definitions $s_{j_1j_2} \equiv \sin\f{(j_1-j_2)\pi}{n}$, $c_{j_1j_2} \equiv \cos\f{(j_1-j_2)\pi}{n}$, and the ones similar to them. Considering the relevant operators in table~\ref{qpnT} and table~\ref{qpnJ}, we get the R\'enyi mutual information organized by the large central charge
\be
I_n^{(0,1)} = I_{n}^\cL + I_{n,(0,1)}^\NL + \cdots.
\ee
Here the leading part $I_{n}^\cL$ equals $I_n^\cl$ (\ref{Incl}) and the next-to-leading part
\be
I_{n,(0,1)}^\NL = I_{n,g}^\oloop + I_{n,v}^\oloop,
\ee
with $I_{n,g}^\oloop$ being (\ref{Ingoloop}) and $I_{n,v}^\oloop$ being (\ref{Involoop}).

\subsection{Contributions of $T$, $J$, $G$ and $H$ in SCFT}

\begin{table}\centering
\begin{tabular}{|c|c|c|c|c|}\cline{1-5}
  level & operators & ? & degeneracy & number \\ \cline{1-5}

  $\f32$ & $G,~H$ & \texttimes & $2n$ & $2n$ \\ \cline{1-5}

  \multirow{2}*{$\f52$} & $\mD_1,~\mD_2$ & \texttimes & $2n$      & \multirow{2}*{$2n^2$} \\ \cline{2-4}
                        & $JG,~JH$       & \texttimes & $2n(n-1)$ & \\ \cline{1-5}

    & $\mE$    & \texttimes & $n$      & \\ \cline{2-4}
  3 & $GH$     & \texttimes & $n(n-1)$ & $n(2n-1)$ \\ \cline{2-4}
    & $GG,~HH$ & \checked   & $n(n-1)$ & \\ \cline{1-5}

         & $\mB,~\mC,~\mF_1,~\mF_2,~\mF_3,~\mF_4$ & \texttimes                & $6n$                     & \\ \cline{2-4}
         & $TG,~TH,~\mM G$                        & \multirow{2}*{\texttimes} & \multirow{2}*{$6n(n-1)$} & \\
  $\f72$ & $\mM H,~J\mD_1,~J\mD_2$                &                           &                          & $n^2(n+5)$ \\ \cline{2-4}
         & $JJG,~JJH$                             & \texttimes                & $n(n-1)(n-2)$            & \\ \cline{2-4}
         & $I(JG),~I(JH)$                         & \texttimes                & $2n(n-1)$                & \\ \cline{1-5}

    & $\mI_1,~\mI_2,~\mI_3,~\mI_4$ & \texttimes                & $4n$                     & \\ \cline{2-4}
    & $J\mE,~G\mD_1,~G\mD_2$       & \multirow{2}*{\texttimes} & \multirow{2}*{$5n(n-1)$} & \\
    & $H\mD_1,~H\mD_2$             &                           &                          & \\ \cline{2-4}
  4 & $JGH$                        & \checked                  & $n(n-1)(n-2)$            & $n (2 n^2+n+1)$ \\ \cline{2-4}
    & $JGG,~JHH$                   & \texttimes                & $n(n-1)(n-2)$            & \\ \cline{2-4}
    & $I(GH)$                      & \texttimes                & $n(n-1)$                 & \\ \cline{2-4}
    & $I(GG),~I(HH)$               & \checked                  & $n(n-1)$                 & \\ \cline{1-5}
\end{tabular}
\caption{Extra holomorphic quasiprimary operators in SCFT$^n$ after further adding contributions of $G$ and $H$. The 5th column agrees with the counting in (\ref{z21}).}
\label{qpnGH}
\end{table}

By further adding the contributions of $G$ and $H$, we have to consider the extra quasiprimary operators that are counted as
\be \label{z21}
(1-x)\chi_{(0)}^n\chi_{(1)}^n(\chi_{(3/2)}^n-1) =  2 n x^{3/2}+2 n^2 x^{5/2}+n (2 n-1) x^3+n^2 (n+5) x^{7/2}+n (2 n^2+n+1) x^4+O(x^{9/2}),
\ee
with $\chi_{(0)}$ being (\ref{chi0}), $\chi_{(1)}$ being (\ref{chi1}) and $\chi_{(3/2)}$ being (\ref{chi32}), and they are listed in table~\ref{qpnGH}.
For the relevant operators we have the normalization factors
\bea
&& \a_{GG}=\a_{HH}=-\f{16c^2}{9}, ~~ \a_{JGH}=\f{16c^3}{27}, ~~ \a_{I(GG)}=\a_{I(HH)}=-\f{32c^2}{3},
\eea
and the OPE coefficients
\be
d_{GG}^{j_1j_2} = - d_{HH}^{j_1j_2} = -\f{3\ii}{32n^3c}\f{1}{s_{j_1j_2}^3}, ~~
d_{JGH}^{j_1j_2j_3} = -\f{9}{64n^4c^2}\f{1}{s_{j_1j_2}s_{j_1j_3}s_{j_2j_3}^2}, ~~
d_{I(GG)}^{j_1j_2} = - d_{I(HH)}^{j_1j_2} =-\f{3\ii}{64n^4c}\f{c_{j_1j_2}}{s_{j_1j_2}^4}.
\ee
Considering the relevant operators in table~\ref{qpnT}, table~\ref{qpnJ}, and table~\ref{qpnGH}, we get the R\'enyi mutual information
\be
I_n^{(0,1,3/2)} = I_{n}^\cL + I_{n,(0,1,3/2)}^\NL + I_{n,(0,1,3/2)}^\NNL \cdots.
\ee
Here the leading part $I_{n}^\cL$ equals $I_n^\cl$ (\ref{Incl}), the next-to-leading part
\be
I_{n,(0,1,3/2)}^\NL = I_{n,g}^\oloop + I_{n,v}^\oloop + I_{n,\psi}^\oloop,
\ee
with $I_{n,g}^\oloop$ being (\ref{Ingoloop}), $I_{n,v}^\oloop$ being (\ref{Involoop}), and $I_{n,v}^\oloop$ being (\ref{Inpsioloop}), and the next-to-next-to-leading part is
\be
I_{n,(0,1,3/2)}^\NNL = \frac{(n+1) (n^2-4) (n^4+40 n^2+679) x^4}{302400 c n^7}+O(x^5).
\ee

\section{R\'enyi entropy of one interval on torus}\label{sec3}

In this section we investigate the R\'enyi entropy of one length $\ell$ interval on a torus with low temperature. The SCFT on torus is just the theory on a circle in thermal state.
We expand the R\'enyi entropy by $q=\exp(-2\pi\b/L)$, with $\b$ being the inverse temperature and $L$ being the spatial period, to order $q^2$. In supergravity side we use the method in \cite{Barrella:2013wja,Chen:2014unl}, and in SCFT side we use the method in \cite{Cardy:2014jwa,Chen:2014unl}.

\subsection{Supergravity result}

The classical part of holographic R\'enyi entropy is \cite{Barrella:2013wja,Chen:2014unl}
\be \label{Sncl}
S_n^\cl = \f{c(n+1)}{6n}\log\Big(\f{L}{\pi \e} \sin\f{\pi\ell}{L}\Big)
        - \Big(\f{c(n-1)(n+1)^2 }{9n^3} \sin^4 \f{\pi\ell}{L}\Big) q^2+O(q^3).
\ee
The one-loop correction to holographic R\'enyi entropy from graviton is \cite{Barrella:2013wja}
\be \label{Sngoloop}
S_{n,g}^\oloop = -\f{2n}{n-1} \bigg( \f{1}{n^4}\f{\sin^4\f{\pi\ell}{L}}{\sin^4\f{\pi\ell}{nL}} -1 \bigg)q^2 + O(q^3).
\ee
For the vector field we get
\bea \label{Snvoloop}
&&\hspace{-1cm}
S_{n,v}^\oloop= -\f{2n}{n-1} \bigg\{ \bigg( \f{1}{n^{2}}\f{\sin^2\f{\pi\ell}{L}}{\sin^2\f{\pi\ell}{nL}} -1\bigg)q
+\bigg[ \f{2\sin^2\f{\pi\ell}{L}}{n^4\sin^2\f{\pi\ell}{nL}}  \bigg(
       n^2\cos^2\f{\pi\ell}{L}
      -n\sin\f{2\pi\ell}{L}\cot\f{\pi\ell}{nL}\\
&&\hspace{-1cm}\phantom{S_{n,\mX}^\oloop=}
      +\f{\sin^2\f{\pi\ell}{L}}{\sin^2\f{\pi\ell}{nL}} \Big(\cos^2\f{\pi\ell}{nL}+\f34 \Big)  \bigg)+\f{\sin^4\f{\pi\ell}{L}}{2n^4}\sum_{j=1}^{n-1} \bigg( \f{1}{\big(\sin^2\f{\pi j}{n}-\sin^2\f{\pi\ell}{nL}\big)^2}
                                                                + \f{1}{\sin^4\f{\pi j}{n}} \bigg)
      -\f{3}{2}\bigg]q^2 + O(q^3) \bigg\}.\nn
\eea
For the two gravitinos we get
\be  \label{Snpsioloop}
S_{n,\psi}^\oloop = -\f{4n}{n-1} \bigg( \f{1}{n^3}\f{\sin^3\f{\pi\ell}{L}}{\sin^3\f{\pi\ell}{nL}} - 1\bigg)q^{3/2} + O(q^{5/2}).
\ee

\subsection{Contributions of $T$ in SCFT}

In this subsection we consider the contributions of only the vacuum conformal family and repeat the calculation in \cite{Cardy:2014jwa,Chen:2014unl}. One has the density matrix in expansion of low temperature
\be \label{rvac}
\r_\vacc = \uum\vac + \f{q^2}{\a_T} |T\rag \lag T| + O(q^3).
\ee
Denoting the interval by $A$ and its complement by $B$, one gets the R\'enyi entropy
\be
S_{n}^{(0)} = -\f{1}{n-1} \log\tr_A\big(\tr_B\uum\vac\big)^n
              -\f{2n}{n-1}\bigg( \f{\tr_A\big[\tr_B|T\rag\lag T|(\tr_B\uum\vac)^{n-1}\big]}
                                  {\a_T\tr_A\big(\tr_B\uum\vac\big)^n} -1 \bigg)q^2 + O(q^3).
\ee
From the state operator correspondence one has
\be
\f{\tr_A\big[\tr_B|T\rag\lag T|(\tr_B\uum\vac)^{n-1}\big]}{\a_T\tr_A\big(\tr_B\uum\vac\big)^n} = \f{\lag T(\inf)T(0)\rag_{\mC^n}}{\a_T}.
\ee
The $n$-fold complex plane $\mC^n$ with coordinate $z$ can be mapped to a complex plane $\mC$ with coordinate $f$ by the conformal transformation
\be
f(z)= \bigg( \f{z-\ep^{\ii\pi\ell/L}}{z-\ep^{-\ii\pi\ell/L}} \bigg)^{1/n}.
\ee
Then one gets
\be
\f{\lag T(\inf)T(0)\rag_{\mC^n}}{\a_T} =  \f{c(n^2-1)^2}{18n^4}\sin^4 \f{\pi\ell}{L}
                                        + \f{1}{n^{4}}\f{\sin^4\f{\pi\ell}{L}}{\sin^4\f{\pi\ell}{nL}}.
\ee
The R\'enyi entropy can be organized by large central charge as
\be
S_n^{(0)} = S_{n,(0)}^\cL + S_{n,(0)}^\NL + \cdots,
\ee
with the leading part $S_{n,(0)}^\cL$ equaling $S_n^\cl$ (\ref{Sncl}) and the next-to-leading part $S_{n,(0)}^\NL$ equaling $S_{n,g}^\oloop$ (\ref{Sngoloop}).

\subsection{Contributions of $J$ in SCFT}

Adding the contributions of $J$, we have to add the density matrix (\ref{rvac}) by
\be \label{rJ}
\r_J=\f{q}{\a_J} |J\rag\lag J| + \f{q^2}{\a_{\mM}}|\mM\rag\lag\mM| + \f{q^2}{\a_{\p J}}|\p J\rag\lag \p J| +O(q^3).
\ee
We get the extra contributions of $J$ to the R\'enyi entropy
\bea
&& S_{n}^{(1)} = -\f{2n}{n-1} \bigg\{
      \bigg(\f{\lag J(\inf)J(0) \rag_{C^n}}{\a_J}-1\bigg)q
    + \bigg[ \f{\lag \mM(\inf)\mM(0) \rag_{\mC^n}}{\a_\mM}
            +\f{\lag \p J(\inf)\p J(0) \rag_{\mC^n}}{\a_{\p J}} \nn\\
&& \phantom{S_{n}^{(1)} =}
            -\f{n}{2}\bigg(\f{\lag J(\inf)J(0) \rag_{\mC^n}}{\a_J}\bigg)^2
            +\f{1}{2} \sum_{j=1}^{n-1} \f{\lag J(\inf)J(0) J(\inf_j)J(0_j) \rag_{\mC^n}}{\a^2_J}-\f{3}{2}
      \bigg]q^2 +O(q^3) \bigg\}.
\eea
With some calculation we get
\bea
&& \f{\lag J(\inf)J(0)\rag_{\mC^n}}{\a_J} = \f{1}{n^2}\f{\sin^2\f{\pi\ell}{L}}{\sin^2\f{\pi\ell}{nL}}, ~~
   \f{\lag \mM(\inf)\mM(0)\rag_{\mC^n}}{\a_\mM} = \f{1}{n^4}\f{\sin^4\f{\pi\ell}{L}}{\sin^4\f{\pi\ell}{nL}}, \nn\\
&& \f{\lag\p J(\inf)\p J(0)\rag_{\mC^n}}{\a_{\p J}}=
     \f{2}{n^4} \f{\sin^2\f{\pi\ell}{L}}{\sin^2\f{\pi\ell}{nL}}
     \bigg( n^2\cos^2\f{\pi\ell}{L}
           -n\sin\f{2\pi\ell}{L}\cot\f{\pi\ell}{nL}
           +\f{\sin^2\f{\pi\ell}{L}}{\sin^2\f{\pi\ell}{nL}} \Big(\cos^2\f{\pi\ell}{nL}+\f12\Big)  \bigg), \nn\\
&&\f{\lag J(\inf) J(0) J(\inf_j)J(0_j)\rag_{\mC^n}}{\a_{J}^2} = \f{\sin^4\f{\pi\ell}{L}}{n^4}
                  \bigg( \f{1}{\big(\sin^2\f{\pi j}{n}-\sin^2\f{\pi\ell}{n L}\big)^2}
                       + \f{1}{\sin^4\f{\pi j}{n}}
                       + \f{1}{\sin^4\f{\pi\ell}{n L}} \bigg).
\eea
Then we reproduce the one-loop gravity result $S_{n,v}^\oloop$ (\ref{Snvoloop}).

\subsection{Contributions of $G$ and $H$ in SCFT}

Adding further the contributions of $G$ and $H$, we need to add the density matrix (\ref{rvac}) with (\ref{rJ}) and
\be
\r_{G,H}=\f{q^{3/2}}{\a_G} |G\rag\lag G| + \f{q^{3/2}}{\a_H} |H\rag\lag H| +O(q^{5/2}).
\ee
We get the extra contribution for R\'enyi entropy from $G$ and $H$
\be
S_n^{(3/2)}=-\f{2 n}{n-1} \bigg(\f{\lag G(\inf)G(0) \rag_{C^n}}{\a_{G}}+\f{\lag H(\inf)H(0) \rag_{C^n}}{\a_{H}}-2\bigg)q^{3/2} +O(q^{5/2}).
\ee
With the correlation functions
\be
\f{\lag G(\inf)G(0) \rag_{C^n}}{\a_{G}}=
\f{\lag H(\inf)H(0) \rag_{C^n}}{\a_{H}}=
\f{1}{n^3}\frac{\sin^3\frac{\pi \ell}{L}}{\sin ^3\frac{\pi \ell}{nL}},
\ee
we reproduce the one-loop gravity result $S_{n,\psi}^\oloop$ (\ref{Snpsioloop}).

\section{Conclusion and discussion}\label{sec4}

In this paper, we have investigated the holographic R\'enyi entropy in the correspondence of $\mN=2$ supergravity under AdS$_3$ background and two-dimensional $\mN=(2,2)$ SCFT. For two short intervals on complex plane with small cross ratio $x \ll 1$ we got the R\'enyi mutual information to order $x^4$, and for one interval on torus with low temperature $q =\ep^{-2\pi\b/L} \ll 1$ we got the R\'enyi entropy to order $q^2$.

Such orders are lower than the previous results in literature, for examples, order $x^8$ for complex plane case in \cite{Chen:2013dxa} and order $q^4$ for torus case in \cite{Chen:2015uia}. This is because in the present case we have the spin-1 field in supergravity side and the current with conformal weight 1 in SCFT side. The small spin in supergravity and small conformal weight in SCFT make the calculation to higher orders more complicated. To get higher order results in supergravity side we need to consider the contributions of $m$-CDW's for the complex plane case and the $m$-letter words in the torus case with $m \geq 3$. In SCFT side the number of degenerate quasiprimary operators would increase very quickly in higher levels, and the orthogonalization of these primary and quasiprimary operators would be very complicated. Although much more complicated, it would be nice if higher order results can be got in the future, and perhaps more effective method has to be developed.

\section*{Acknowledgements}

The authors would like to thank Jun-Bao Wu and Hao Ouyang for valuable discussions and suggestions. And also we special thank Matthew Headrick for his Mathematica code \emph{Virasoro.nb} which could be downloaded at the personal homepage \url{http://people.brandeis.edu/~headrick/Mathematica/index.html}.
ZL was supported by NSFC Grant No.~11222549 and No.~11575202.

\appendix

\section{Review of two-dimensional $\mN$=(2,2) SCFT}\label{appA}

In this appendix we review some useful properties of the $\mN=(2,2)$ SCFT. More details can be found in \cite{Blumenhagen:2009zz}.

In the $\mN=(2,2)$ SCFT, we have the stress tensor $T(z)$, $\bar T(\bar z)$, the current $J(z)$, $\bar J(\bar z)$, and their supersymmetric partners $G(z)$, $\bar G(\bar z)$, $H(z)$, $\bar H(\bar z)$, as well as the operators that are constructed from them. These operators can be written as the products of the holomorphic ones and the anti-holomorphic ones, and there is one-to-one correspondence between the holomorphic and anti-holomorphic operators. So in this paper we only need to consider the holomorphic sector of the SCFT.

With only $T$, we get the character of the vacuum conformal family
\be \label{chi0}
\chi_{(0)} = \prod_{k=0}^\inf \f{1}{1-x^{k+2}}.
\ee
The modes of $T$ are denoted by $L_m$ with $m\in\Z$, and they form the Virasoro algebra
\be \label{LmLn}
[L_m, L_n]=(m-n) L_{m+n}+\f{c}{12} (m^{3}-m) \d_{m+n}.
\ee
The primary operator of the vacuum conformal family is the identity operator. At level 2 we have $T$ with normalization factor $\a_T=\f{c}{2}$, and at level 4 we have quasiprimary operator
\be
\mA=(TT)-\f{3}{10}\p^2T, ~~ \a_{\mc A}=\f{c(5c+22)}{10}.
\ee
Here we use the bracket `()' to denote normal ordering.
Under a general conformal transformation $z \to f(z)$, they transform as
\be
T(z)=f'^2 T(f)+\f{c}{12}s, ~~
\mA(z)=f'^4\mA(f)+\f{5c+22}{30}s \Big( f'^2 T(f)+\f{c}{24}s \Big),
\ee
with the Schwarzian derivative being
\be
s(z)=\f{f'''(z)}{f'(z)}-\f32 \bigg( \f{f''(z)}{f'(z)} \bigg)^2.
\ee

By adding the current operator $J$ with conformal weights $(1,0)$ we get the bosonic part of the SCFT\footnote{The term `bosonic part' is not so precise, since with the fermionic operators $G$ and $H$ we can also construct bosonic operators. Here by bosonic part we mean the sector with operators that are constructed solely by $T$ and $J$, without $G$ or $H$.}, and we have to multiply the character (\ref{chi0}) by
\be \label{chi1}
\chi_{(1)} = \prod_{k=0}^\inf \f{1}{1-x^{k+1}}.
\ee
The Virasoro algebra (\ref{LmLn}) is supplemented by
\be \label{LmJnJmJn}
[L_m, J_n] = -n J_{m+n}, ~~ [J_m, J_n]=\f{c}{3} m \d_{m+n}.
\ee
Note that the algebra of SCFT bosonic part is close and self-consistent. Besides the ones in vacuum conformal block, there are other quasiprimary operators that can be counted as
\be
(1-x)\chi_{(0)}(\chi_{(1)}-1) = x + x^2 + 2x^3 + 4x^4 + O(x^5).
\ee
Among these quasiprimary operators, the primary ones can be counted as
\be
\f{\chi_{(0)}(\chi_{(1)}-1)}{\chi} = x + x^2 + x^3 + 2x^4 + O(x^5), ~~ \chi \equiv \prod_{k=1}^\inf \f{1}{1-x^{k}}.
\ee
These primary and quasiprimary operators are listed in table~\ref{qpJ}. For the primary operator $J$ we have the normalization $\a_J=\f{c}{3}$, for quasiprimary operator $\mK$ we have
\be
\mK=(TJ)-\f{1}{2}\p^2J, ~~
\a_\mK=\f{c(c+2)}{6},
\ee
and for primary operator $\mM$ we have
\be
\mM = (JJ) - \f23 T, ~~ \a_\mM = \f{2c(c-1)}{9}.
\ee
Under a conformal transformation $z \to f(z)$ we have
\be
J(z) = f' J(f), ~~
\mK(z) = f'^3\mK(f) + \f{c+2}{12}s f' J(f).
\ee
In this paper we need the structure constants
\be
C_{JJJ}=0, ~~ C_{TJJ}=\frac{c}{3}, ~~ C_{JJ\mM}=\frac{2c(c-1)}{9},
\ee
as well as the four-point function on complex plane
\be
\lag J(z_1)J(z_2)J(z_3)J(z_4) \rag_\mC = \frac{c^2 }{9} \Big(  \frac{1}{z_{12}^2 z_{34}^2}
                                                          +\frac{1}{z_{13}^2 z_{24}^2}
                                                          +\frac{1}{z_{14}^2 z_{23}^2} \Big),
\ee
with the shorthand $z_{jk}=z_j-z_k$.

\begin{table}
  \centering
\begin{tabular}{|c|c|c|c|c|} \hline
  level         & 1   &  2    &  3    & 4             \\ \hline
  $J$           & $J$ &       & $\mK$ & $\mL$         \\ \hline
  $\mM$         &     & $\mM$ &       & $\mN$         \\ \hline
  $\mO$         &     &       & $\mO$ &               \\ \hline
  $\mP_1,\mP_2$ &     &       &       & $\mP_1,\mP_2$ \\ \hline
\end{tabular}
\caption{Up to level 4 the extra quasiprimary operators because of the adding of $J$. In the first column we list the primary operator of each conformal family.}\label{qpJ}
\end{table}

\begin{table}
  \centering
\begin{tabular}{|c|c|c|c|c|c|} \hline
  level                        & 3/2           &  5/2          &  3    & 7/2                       & 4         \\ \hline
  $G$                          & $G$           &               &       & $\mB$                     &           \\ \hline
  $H$                          & $H$           &               &       & $\mC$                     &           \\ \hline
  $\mD_1,\mD_2$                &               & $\mD_1,\mD_2$ &       &                           &           \\ \hline
  $\mE$                        &               &               & $\mE$ &                           &           \\ \hline
  $\mF_1,\mF_2,\mF_3,\mF_4$    &               &               &       & $\mF_1,\mF_2,\mF_3,\mF_4$ &           \\ \hline
  $\mI_1,\mI_2,\mI_3,\mI_4$    &               &               &       &                           & $\mI_1,\mI_2,\mI_3,\mI_4$ \\ \hline
\end{tabular}
\caption{The extra primary and quasiprimary operators because of the adding of $G$ and $H$.}\label{qpGH}
\end{table}

By further adding the fermionic operators $G$ and $H$ with conformal weights $(3/2,0)$, we get the SCFT and the character is the product of (\ref{chi0}), (\ref{chi1}) and
\be \label{chi32}
\chi_{(3/2)} = \prod_{k=0}^\inf \big(1 + x^{k+3/2}\big)^2.
\ee
Note that there are two primary operators at level $3/2$ and so there is a power 2 in the right-hand side of the above equation. As argued in \cite{Zhang:2015hoa}, in large central charge limit we only need to consider the Neveu-Schwarz sector for the fermionic operators. We have the modes $G_r, H_s$ with $r, s \in \Z+1/2$, and to compose the $\mN=2$ super Virasoro algebra we need (\ref{LmLn}), (\ref{LmJnJmJn}), and
\bea
&& [L_m, G_r]=(\f{m}{2}-r) {G}_{m+r} , ~~
   [L_m, {H}_r]=(\f{m}{2}-r) {H}_{m+r},\nn\\
&& [J_m,{G}_r]={H}_{m+r}  , ~~
   [J_m, {H}_r]= {G}_{m+r} , ~~
   \{{G}_r, {H}_s \}=-2(r-s)J_{r+s}, \\
&& \{{G}_r, {G}_s \}= 4L_{r+s}+\f{2c}{3} (r^{2}-\f{1}{4}) \d_{r+s}, ~~
   \{{H}_r, {H}_s \}=-4L_{r+s}-\f{2c}{3} (r^{2}-\f{1}{4}) \d_{r+s}.  \nn
\eea
The extra quasiprimary operators can be counted as
\be
(1-x)\chi_{(0)}\chi_{(1)}(\chi_{(3/2)}-1) =  2 x^{3/2}+2 x^{5/2}+x^3+6 x^{7/2}+4 x^4 + O(x^{9/2}),
\ee
among which the primary ones can be counted as
\be
\f{\chi_{(0)}\chi_{(1)}(\chi_{(3/2)}-1)}{\chi} = 2 x^{3/2}+2 x^{5/2}+x^3+4 x^{7/2}+4 x^4 + O(x^{9/2}).
\ee
These operators are listed in table~\ref{qpGH}. For $G$ and $H$ we have normalization constants $\a_G=-\a_H=\f{4c}3$, and under conformal transformation they transform as
\be
G(z)=f'^{\f{3}{2}} G(f),  ~~  H(z)=f'^{\f{3}{2}} H(f).
\ee
There are useful structure constants
\be
C_{JJG}=C_{JJH}=C_{JGG}=C_{JHH}=0, ~~ C_{JGH}=-\frac{4 c}{3}.
\ee

\section{Some useful summation formulas} \label{appB}

In this appendix we collect some summation formulas that are useful to calculation in section~\ref{sec2}. We define
\be
f_m=\sum_{j=1}^{n-1}\f{1}{ \lt( \sin\f{\pi j}{n} \rt)^{2m}},
\ee
and we need
\bea
&& f_2=\frac{(n^2-1) \left(n^2+11\right)}{45} , ~~~
   f_3=\frac{(n^2-1)  \left(2 n^4+23 n^2+191\right)}{945} ,    \nn\\
&& f_4=\frac{(n^2-1) \left(n^2+11\right) \left(3 n^4+10 n^2+227\right)}{14175}.
\eea
We have
\bea
&& \sum_{\neq} \f{1}{s_{j_1j_2}^4 s_{j_1j_3}^4} = \frac{4 n (n^2-1) (n^2-4)(n^2+11)(n^2+19)}{14175}, \nn\\
&& \sum_{\neq} \f{1}{s_{j_1j_2}^2 s_{j_1j_3}^2 s_{j_2j_3}^2} = \frac{2 n (n^2-1) (n^2-4) (n^2+47)}{945}, \\
&& \sum_{\neq} \f{1}{s_{j_1j_2}^2 s_{j_1j_3}^2 s_{j_2j_3}^4} = \frac{2 n (n^2-1) (n^2-4)(n^4+40 n^2+679)}{14175}, \nn
\eea
with definition $s_{j_1j_2}\equiv\sin\f{\pi(j_1-j_2)}{n}$ and the ones similar to it, and the summation is over $0 \leq j_{1,2,3} \leq n-1$ with constraints $j_1\neq j_2$, $j_1\neq j_3$, $j_2\neq j_3$. We have
\bea
&& \sum_{\neq} \f{1}{s_{j_1j_2}^4 s_{j_3j_4}^4}  = \frac{n (n^2-1) (n-2)(n-3) (n^2+11) (7 n^3+13 n^2+93 n+127)}{14175} , \\
&& \sum_{\neq} \f{1}{s_{j_1j_2}^2 s_{j_3j_4}^2 s_{j_1j_3}^2 s_{j_2j_4}^2} = \frac{4 n (n^2-1) (n^2-4)(n^2-9) (n^2+119)}{14175}, \nn
\eea
with the summation being over $0 \leq j_{1,2,3,4} \leq n-1$ and the constraints being $j_1\neq j_2$, $j_1\neq j_3$, $j_1\neq j_4$, $j_2\neq j_3$,  $j_2\neq j_4$, $j_3\neq j_4$. Finally we have
\be
\sum{'} \f{1}{s_{j_1j_2}^2 s_{j_2j_3}^2 s_{j_3j_4}^2 s_{j_4j_1}^2} = \frac{4 n (n^2-1) (n^2-4) (3 n^4+170 n^2-653)}{14175},
\ee
with the summation being over $0 \leq j_{1,2,3,4} \leq n-1$ and the constraints being
\be
j_1\neq j_2, ~ j_2\neq j_3, ~  j_3\neq j_4, ~ j_4\neq j_1, ~ (j_1,j_2) \neq (j_3,j_4).
\ee

\providecommand{\href}[2]{#2}\begingroup\raggedright\endgroup


\end{document}